\documentclass[a4paper]{jpconf}
\usepackage{graphicx}

\newcommand{\Trace}{\mathrm{Tr\hspace{0.5pt}}}
\newcommand{\eq}{\begin{equation}}
\newcommand{\qe}{\end{equation}}
\newcommand{\de}{\mathrm{d}\hspace{1pt}}
\newcommand{\eqa}{\begin{eqnarray}}
\newcommand{\qea}{\end{eqnarray}}
\newcommand{\De}{\mathcal{D}\hspace{1pt}}
\newcommand{\avg}[1]{{\langle{#1}\rangle}}
\newcommand{\Real}{\mathrm{Re}\hspace{1pt}}

\begin{document}

\title{Phase transitions in heavy-quark QCD \\ from an effective theory}

\author{M.~Fromm, J.~Langelage\footnote{Current address: Institut f\"ur theoretische Physik, ETH, Wolfgang-Pauli-Str.~27, 8049 Z\"urich, Switzerland.}, S.~Lottini\footnote{Speaker at the conference.}$^,$\footnote{Current address: NIC, DESY, Platanenallee 6, 15738 Zeuthen, Germany.}, M.~Neuman, O.~Philipsen}

\address{ITP, Goethe-Universit\"at, Max-von-Laue-str.~1, 60438 Frankfurt am Main, Germany}

\ead{lottini@th.physik.uni-frankfurt.de}

\begin{abstract}
	With combined hopping parameter and strong coupling expansions, we calculate a dimensionally reduced 
	Polyakov-loop effective theory valid for heavy quarks at nonzero temperature and arbitrary chemical 
	potential. We numerically compute the critical endpoint of the deconfinement transition as a function 
	of quark masses and number of flavours. We also investigate the applicability of the model to the low-T
	and high density region, specifically in terms of baryon condensation phenomena.
\end{abstract}

\section{Introduction}
Despite its brilliant achievements in tackling perturbative problems, analytical investigations of QCD can not do much when it comes
to determining the structure of its phase diagram in the $(T,\mu)$ plane, with $T$ temperature and $\mu$ quark chemical potential.
In nonpertubative regimes, the answers numerical lattice QCD calculations can provide are mainly limited to $\mu\simeq 0$: the notoriuos
sign problem renders the traditional Monte Carlo sampling approach meaningless and one has to rely on alternative methods, whose efficiency
typically degrades about $\mu\sim T$: examples are using an imaginary chemical potential, performing a Taylor expansion around $\mu=0$ and
reweighting of ensembles generated at zero chemical potential.

Yet, the relevant physics for heavy-ion collisions, as well as for compact-star astrophysics, takes place at far-from-zero $\mu$:
it would then be desirable to devise strategies able somehow to circumvent the problems. In this category fall the many effective
approaches developed throughout the years, such as sigma models, (p)NJL models and so on.

In particular, in characterising the $T\to 0$ region of the $(T,\mu)$ phase diagram, the phenomenon of \textit{Silver Blaze} is expected
based on very general considerations: there, the partition function -- hence all interesting observables -- have to be independent of
the chemical potential, up to a critical value associated to the lightest baryonic state in the theory.
In the path-integral representation of QCD, the Dirac eigenvalues display an explicit dependence on $\mu$, therefore the Silver Blaze phenomenon
requires a highly non-trivial cancellation between them.
Due to the shortcomings of the numerical methods mentioned above, the phenomenon has not been reproduced successfully in lattice QCD so far;
also on the analytical side, an explicit proof exists only for the choice of an isospin chemical potential 
$\mu_I = \mu_u = -\mu_d$, with a critical onset located at $\mu_I=m_\pi/2$ \cite{cohen_silverblaze}.

It is perhaps possible to avoid the sign problem altogether, thus exploring numerically the cold dense region $\mu \ll T$ of QCD, by
employing the so-called Langevin dynamics \cite{langevin}: recently, numerical results for a scalar complex field have been obtained \cite{aarts_silverblaze}
and later reproduced in the context of a flux representation for the same theory \cite{gattringer_silverblaze};
both techniques, however, are at the moment hardly applicable to QCD, and it is not at all guaranteed that they will ever be.

In this work we present an effective, dimensionally-reduced Polyakov-loop model obtained by applying systematically
strong-coupling and hopping-parameter expansions to the original QCD lattice formulation, valid as long as quarks are not too light
and the gauge coupling $\beta$ is not too large;
after briefly describing the theory (Section \ref{sec:efftheory}),
we apply it to the cold dense region (Section \ref{sec:silverblaze}) and show that the signature of the Silver Blaze phenomenon can be 
read off the numerical results.
A more detailed presentation of the theory can be found in \cite{efftheo_gauge, efftheo_mu};
for the Silver Blaze property in this context, see also \cite{efftheo_silver}.

\section{The effective theory}
\label{sec:efftheory}
The model lives on a three-dimensional lattice, with complex scalars as per-site degrees of freedom, representing
the traced Polyakov loops $L_x\equiv \Trace(W_x), W_x\in SU(3)$.
The associated effective action consists of a pure-gauge term and a fermionic contribution; the former comes from
application of strong-coupling methods to the original four-dimensional Yang-Mills gauge action, and the latter is
obtained with a hopping-parameter expansion, starting from the standard Wilson expression for the fermionic
term in lattice QCD. In general, then, we have
\eq
	Z_\mathrm{eff}(\ldots) = \int \underbrace{\Big( \prod_x \de L_x e^{V(L_x)} \Big)}_{[\De L]} e^{-S_\mathrm{eff}[L]} \;\;,\;\;
		 \int \de L e^{V(L)} = \int_{-\pi}^{+\pi}\de \theta \int_{-\pi}^{+\pi}\de \phi \hspace{5pt} e^{2V\big(L(\theta,\phi)\big)}\;\;,
	\label{eq:z_eff_general} 
\qe
where the parametrisation of the scalar $L$ is made explicit (see \cite{efftheo_gauge}).
The partition function $Z$ depends on effective couplings which are in turn some functions of the original couplings 
appearing in the four-dimensional lattice QCD action.

Let us now sketch how the pure-gauge part of the effective action is obtained \cite{efftheo_gauge} starting from
the four-dimensional Euclidean Yang-Mills partition function 
at finite temperature (with time extent $N_\tau$ lattice spacings) 
\eq
	Z_{SU(3),4d} = \int[\mathcal{D}U_\ell] \exp\Big\{ \frac{\beta}{2N}
		\sum_{p} (\Trace U_{p}+\Trace U_{p}^\dagger) \Big\}
		\;,\;\; \beta=\frac{2N}{g^2} = 18 u + \ldots \;\;.
	\label{eq:original_ym}
\qe
By integrating out the spatial degrees of freedom
and subsequently applying a strong-coupling expansion, 
one gets a tower of interactions in the effective theory,
the leading one of which is simply a ``spin-spin'' term
connecting nearest neighbours in the fundamental representation (the superscript ``$s$'' denotes the 
compliance of the terms to the centre symmetry requirement):
\eq
	-S_\mathrm{eff}^s = \lambda_1 S_1^s + \lambda_2 S_2^s \cdots
		\;;\;
	\lambda_1 S_1^s = \lambda_1(\beta, N_\tau)\sum_{\avg{ij}}(L_iL_j^*+L_i^*L_j)
		\;;\;
	\lambda_{i>1} = o(\lambda_1)~\mbox{for}~\beta\to 0\;.
\qe
The precise form of the map $\lambda_1(\beta, N_\tau)$ comes from an order-by-order
enumeration of strong-coupling graphs having the two Polyakov lines as boundary: we calculated 
the series up to order $u^{10}$ \cite{efftheo_gauge}.\footnote{When fermions are introduced, a shift in the gauge maps between couplings is induced \cite{efftheo_mu}.}
Moreover, one can resum higher powers of (certain classes of) graphs and improve the effective theory to
a ``nonlinear'' form
\eq
	Z_\mathrm{eff} = \int [\De L] \prod_\avg{ij}(1+2\lambda_1 \Real L_i L_j^*)\;\;; 
\qe
the important point is that this model reproduces the deconfinement transition of the original
theory, Eq.~(\ref{eq:original_ym}), and once the critical point (that still falls within the range
of applicability of the strong-coupling methods) is known, it can be translated back to
a table $\beta_c(N_\tau)$ with a sufficient precision and in a wide enough range of $N_\tau$ 
to allow for a sensible continuum extrapolation of the physical deconfinement point $T_c$ \cite{efftheo_gauge}.

The next step is the inclusion of fermions in the theory: we perform an expansion in the hopping parameter $\kappa$,
therefore the masses (we consider $N_f$ degenerate quarks of mass $M = (1-8\kappa)/(2a\kappa)$) must be large enough to guarantee convergence.
The Wilson action for quarks is then written as
\eq
	-S_q = -N_f \sum_{\ell=1}^\infty \frac{\kappa^\ell}{\ell} \Trace (H[U]^\ell)
\qe
with $H$ is the hopping matrix. Its structure is such that the non-zero contributions after integration of the gauge fields
are given by various kinds of closed loops of length $\ell$; each of their links carries a factor $\kappa$
(and, if the chemical potential is turned on, an additional factor $e^{\pm a\mu}$ in the temporal direction):
the expression above is then also an expansion in powers of $\kappa$. The general form of the fermion contribution
to the effective action is analogous to an external field in a spin system (here $a$ denotes the explicit breaking
of centre symmetry):\footnote{The next term seems an ordinary nearest-neighbour
	interaction, but is actually non-centre-symmetric, Eq.~(\ref{eq:efftheo_full}).}
\eq
	-S_\mathrm{eff}^a = -2 N_f \sum_{i=1}^\infty \Big(
		h_i(u,\kappa,\mu,N_\tau)S_i^a + \overline{h}_i(u,\kappa,\mu,N_\tau)S_i^{a,\dagger}
	\Big) \;\; ; \;\;\overline{h}_i(-\mu)=h_i(+\mu)\;\;,
\qe
with the $i$-order reflecting increasingly subdominant terms.
A closer inspection of the leading term $S_1^a$ reveals that, once again, a partial resummation of higher powers 
of the same graphs is possible: one finds that
\eq
	-S_\mathrm{eff}^a = \log\Big[\prod_x
		\det\Big(1 + h_1 W_x\Big)^{2N_f}
		\det\Big(1 + \overline{h}_1 W^\dagger_x\Big)^{2N_f}\Big] + \cdots
	\;\;;\;\; h_1=(2\kappa e^{a\mu})^{N_\tau}(1+\cdots)\;,
\qe

The expression for $h_1$ received also contributions from various types of higher-order graphs; including these, and the first 
subleading fermionic term $S_2^a$, one can write the model as \cite{efftheo_mu, efftheo_silver}:
\eqa
	Z_\mathrm{eff} &=& 
		\int [\De L]
		\prod_{<ij>}\Big(1+2\lambda \Real L_i L_j^*\Big)
		\prod_x \det\Big[(1+h_1 W_x)(1+\overline{h}_1 W^\dagger_x)\Big]^{2N_f} \nonumber \\
		& &\hspace{-0.5cm} \prod_{<ij>} \Big[ 1 - h_2 \Trace\frac{W_i}{1+CW_i} \Trace\frac{W_j}{1+CW_j} \Big]
			\Big[ 1 - \overline{h}_2 \Trace\frac{W^\dagger_i}{1+\overline{C}W^\dagger_i}
				\Trace\frac{W^\dagger_j}{1+\overline{C}W_j^\dagger} \Big] \;\;,
	\label{eq:efftheo_full}
\qea
(we rename $\lambda_1\to\lambda$ and ignore higher-order corrections).
Despite the appearance of $W_x$ and $W_x^\dagger$ in the above, $Z_\mathrm{eff}$
can still be expressed, in practice, in terms only of the $\theta,\phi$ parameterising $L_x$ in Eq.~(\ref{eq:z_eff_general}).
The effective couplings are given by:
\eqa
	h_1 &=& C\Big( 1 + 6\kappa^2 N_\tau \frac{u-u^{N_\tau}}{1-u} + \cdots \Big) \;\; ; \\
	h_2 &=& C^2 \frac{\kappa^2}{3}\Big( 1 + 2\frac{u-u^{N_\tau}}{1-u} \Big) \;\; ; \\
	C   &=& (2\kappa e^{a\mu})^{N_\tau} \;\; .
\qea
The meson and baryon masses are also evaluated:
\eq
	am_M = -2\ln(2\kappa)-6\kappa^2-24\kappa^2\frac{u}{1-u}+\ldots\;;\;\;
	am_B = -3\ln (2\kappa)-18\kappa^2\frac{u}{1-u}+\ldots\;.
\qe

The model, as formulated in Eq.~(\ref{eq:efftheo_full}), is well suited to different simulational strategies: we 
could confirm that the standard Metropolis approach, a flux-representation-based worm algorithm, and a complex-Langevin implementation
all agree with each other: however, one has to keep in mind that the first has to rely on a reweighting procedure for $\mu>0$ 
(which does not hinder its scope significantly for not too large spatial volumes)
and the second ceases to be of use if the $h_2$ term is included, while the third does not have a sign problem at all
and is thus the best choice to deal with large-volume, finite-$\mu$ ensembles. 
All of these algorithms, when compared to traditional lattice QCD approaches,
are far less expensive in terms of required system resources and CPU time. 

\begin{figure}
	\begin{center}
	\raisebox{0.022\textwidth}{\includegraphics[width=0.36\textwidth]{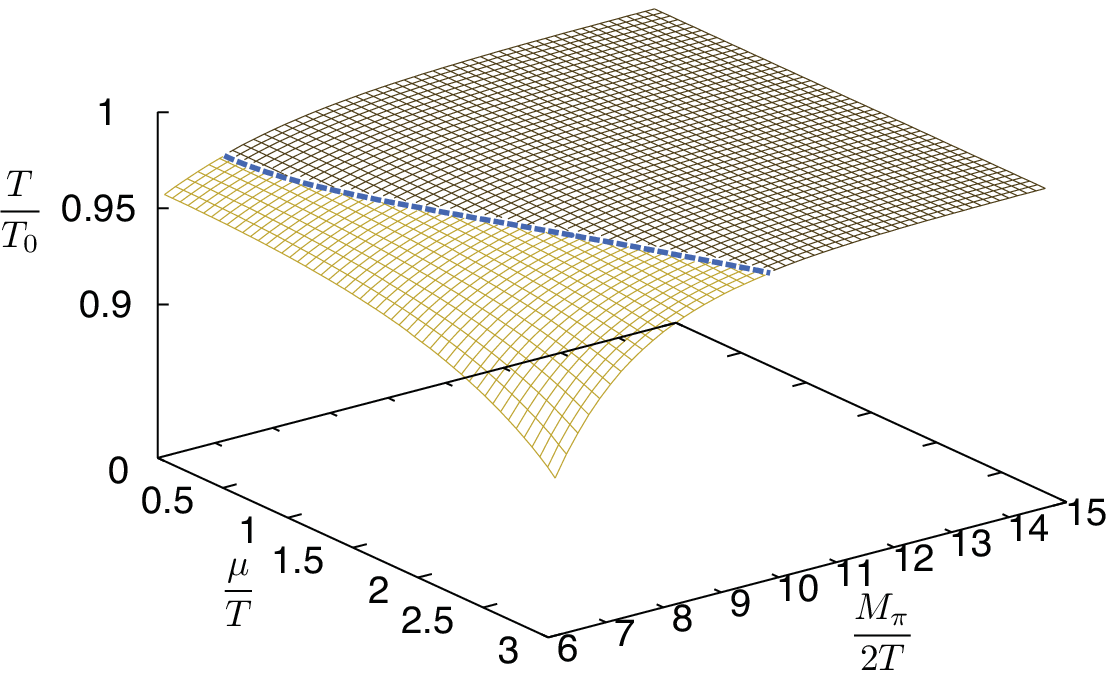}}
	\hspace{0.06\textwidth}
	\includegraphics[width=0.36\textwidth]{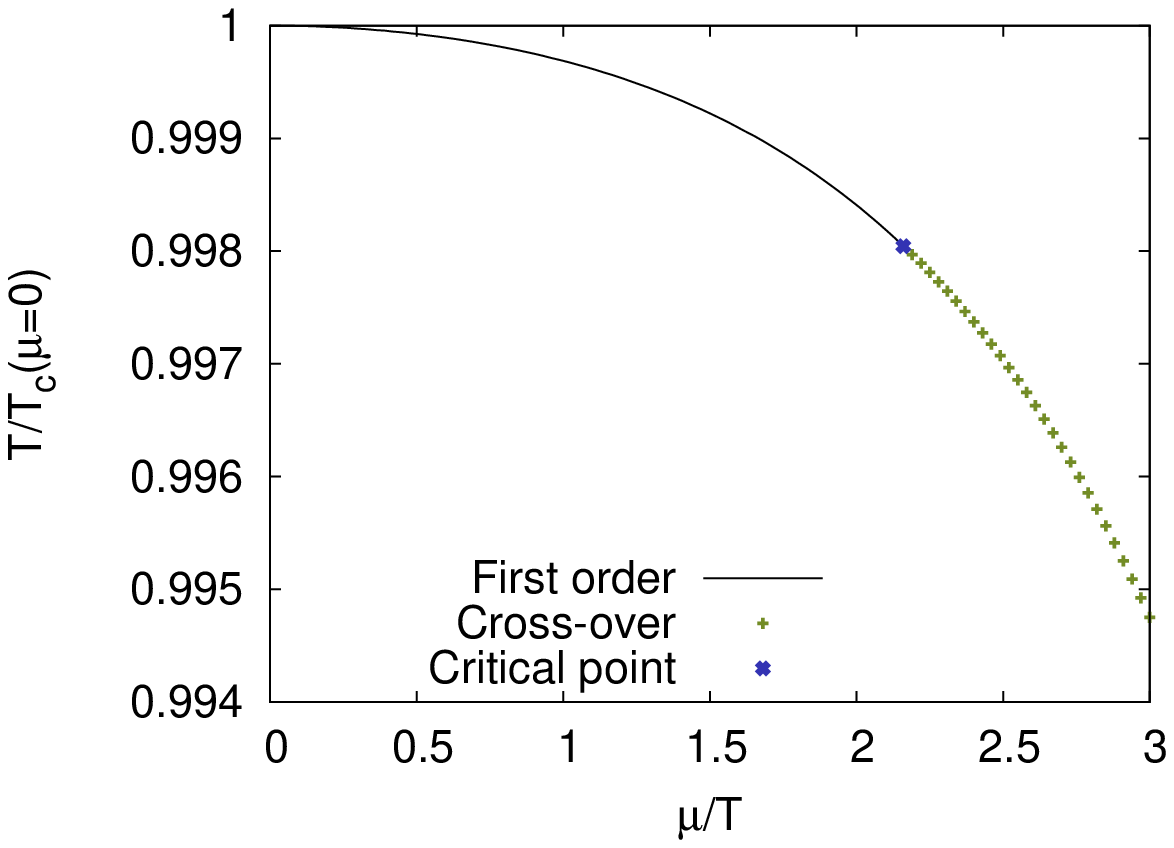}
	\caption{\textit{Left:} $N_f=2$ phase diagram in the space of temperature, chemical potential and meson mass (here called $M_\pi$).
		$T_0$ is the pure-gauge transition point; the blue line is the critical line.
		\textit{Right:} $T$-$\mu$ phase diagram for the choice $N_f=2$, $M_{ud}/T\simeq 8.68$; the vertical axis is normalised to the $\mu=0$
		transition temperature for this mass choice.}
	\label{fig:phasediagram}
	\end{center}
\end{figure}

An important application of this effective approach concerns the mapping of the critical surface in the
$(M_{ud},M_s,\mu/T)$ space. Due to the heavy-quark validity of the model, the second-order surface that could
be located is the one associated to the upper-right corner of the Columbia plot: there, a nice agreement between the
findings at zero, real positive and imaginary $\mu$
was verified and the correct universality class of the transition was confirmed.
Moreover, the result at zero chemical potential for $N_f=1,2,3$
closely reproduces those found in ordinary QCD simulations e.g.~in \cite{deforcrand98, saito11}.
The shape of the critical surface at all chemical potentials can be summarised in the parametrisation \cite{efftheo_mu}:
\eq
	(N_f h_1 e^{-\mu/T})_\mathrm{crit} = \frac{0.00075(1)}{\cosh \mu/T}\;\;;\;\; \kappa_f \simeq e^{-aM_f}/2\;,\;M_f\to\infty\;\;;
\qe
the phase diagram for heavy-quark QCD can then be determined in $(\frac{\mu}{T}, \frac{M_\pi}{2T}, \frac{T}{T_0})$, 
and, by taking slices of it, a ``heavy fermion'' version of the familiar $\mu$-$T$
phase diagram can be drawn (Fig.~\ref{fig:phasediagram}).

\section{Low-temperature regime}
\label{sec:silverblaze}
We now turn to investigate the cold and dense regime of QCD with this effective theory.
In the (analytically solvable) static strong-coupling limit, one finds
that the Silver Blaze property holds as $T \to 0$, with the quark number density approaching
a step function centred at $3\mu=m_B=-3\log(2\kappa)$ (the last equality is exact in this limit):
\eq
	a^3 n \stackrel{T\to0}{\longrightarrow} 2 N_c \hspace{2.5pt} \Theta(\mu_B - m_B
)\;\;;
\qe
one sees then also that a saturation phenomenon is expected
(this is connected to the resummations performed in deriving the form of $Z_\mathrm{eff}$).
The next step is to perform numerical computations with the full expression Eq.~(\ref{eq:efftheo_full}).
Taking advantage of the fact that $N_\tau$ is now merely a parameter in the effective couplings' maps,
we can tune it to very large values in the hundreds: this effectively realises $\lambda\to 0$ thus simplifying the
model; by keeping $\kappa \sim 10^{-3}$, moreover, we can still trust the pure-Yang-Mills
scale-setting prescription \cite{sommer_scale}.
In this way, we can measure numerically the baryon density as a function of the baryon chemical potential
with $m_B=30$ GeV and four different temperatures from $20$ MeV to $2.5$ MeV \cite{efftheo_silver}.

The baryon density $n_B$ can be expressed as a meaningful physical quantity after a careful
continuum limit is taken: we could generate data at nine different values of the
lattice spacings $a$ and perform a fit to the continuum limit;
$n_b$ is measured as
\eq
	a^3 n_b = \frac{a^3}{3} \Big( \frac{T}{V} \frac{\partial \log Z}{\partial \mu} \Big)
		=
		 - \frac{1}{3 N_s^3} \Big\langle \frac{\partial S_\mathrm{eff} }{\partial (\mu/T) } \Big\rangle\;\;. 
\qe 

\begin{figure}
\begin{center}
	\includegraphics[width=0.4\textwidth]{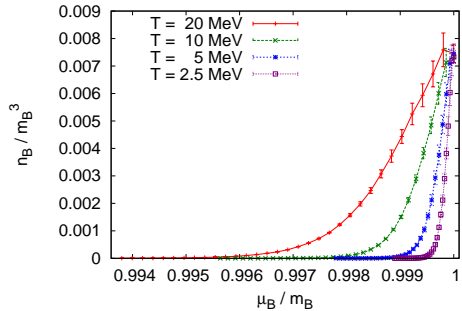}
	\caption{Continuum-extrapolated baryon density, in units of $m_B^3$, as a function of the baryon chemical potential 
		(also measured in units of $m_B$) for various temperatures.}
	\label{fig:silver}
\end{center}
\end{figure}

The results are shown in Fig.~\ref{fig:silver}: the step-like shape of the density is less and less 
smoothed-out as $T\to 0$ (and anyway the $\mu$-scale is extremely tiny); besides, as the baryonic
chemical potential hits the baryon mass the expected saturation to a common, roughly $T$-independent density is seen, and
this value, when expressed in appropriate units of $m_B$, is less than a factor two off the real-world
nuclear matter, despite this model being limited to extremely massive quarks.
This seems to suggest that the core features of nuclear condensation might be largely independent of 
whether the fundamental constituents of the baryon are heavy or light.

\end{document}